# Large Thermoelectric Power Factor in Whisker Crystals of Solid Solutions of the One-Dimensional Tellurides Ta$_4$SiTe$_4$ and Nb$_4$SiTe$_4$

Yuma Yoshikawa, Taichi Wada, Yoshihiko Okamoto*, Yasuhiro Abe, and Koshi Takenaka

*Department of Applied Physics, Nagoya University, Nagoya 464-8603, Japan*
*E-mail: yokamoto@nuap.nagoya-u.ac.jp

One-dimensional tellurides Ta$_4$SiTe$_4$ and Nb$_4$SiTe$_4$ were found to show high thermoelectric performance below room temperature. This study reported the synthesis and thermoelectric properties of whisker crystals of Ta$_4$SiTe$_4$-Nb$_4$SiTe$_4$ solid solutions and Mo- or Ti-doped (Ta$_{0.5}$Nb$_{0.5}$)$_4$SiTe$_4$. Thermoelectric power of the solid solutions systematically increased with increasing Ta content, while their electrical resistivity was unexpectedly small. Mo- and Ti-doped (Ta$_{0.5}$Nb$_{0.5}$)$_4$SiTe$_4$ showed $n$- and $p$-type thermoelectric properties with large power factors exceeding 40 μW cm$^{-1}$ K$^{-2}$, respectively. The fact that not only Ta$_4$SiTe$_4$ and Nb$_4$SiTe$_4$ but also their solid solutions showed high performance indicated that this system is a promising candidate for thermoelectric applications at low temperatures.

Undoped and Mo-doped Ta$_4$SiTe$_4$ have recently been reported to show excellent thermoelectric properties at low temperatures.[1] Ta$_4$SiTe$_4$ has a strongly one-dimensional crystal structure comprising Ta$_4$SiTe$_4$ chains along the $c$ axis of its orthorhombic unit cell with the *Pbam* space group, which are weakly bonded by van der Waals interactions between Te atoms.[2,3] Reflecting the one-dimensional crystal structure, the prepared samples have a whisker form grown along the Ta$_4$SiTe$_4$ chains. The electrical resistivity ρ and thermoelectric power $S$ measured along the whiskers, i.e., parallel to the Ta$_4$SiTe$_4$ chains, indicated that the power factor $P = S^2/\rho$ of undoped and Mo-doped Ta$_4$SiTe$_4$ is significantly larger than those of practical thermoelectric materials below room temperature. Undoped Ta$_4$SiTe$_4$ showed a very large and negative $S$ of ~ −400 μV K$^{-1}$, while maintaining a small ρ of ~ 2 mΩ cm at 100–200 K, yielding a maximum $P$ of 80 μW cm$^{-1}$ K$^{-2}$ at 130 K. The optimum temperature was controlled over a wide range by Mo substitution to the Ta site and the 0.1–0.2% substitution gives rise to a very large $P$ of 170 μW cm$^{-1}$ K$^{-2}$ at 220–280 K,[1] which is almost four times larger than those of Bi$_2$Te$_3$-based materials at room temperature.[4,5]

Mo-doped Nb$_4$SiTe$_4$ also showed high thermoelectric performance as an $n$-type material.[6] Nb$_4$SiTe$_4$ is isostructural and isoelectronic to Ta$_4$SiTe$_4$.[7] The maximum $P$ of Mo-doped Nb$_4$SiTe$_4$ reaches 70 μW cm$^{-1}$ K$^{-2}$ at 230–300 K.[6] Moreover, Ti-doped Ta$_4$SiTe$_4$ and Nb$_4$SiTe$_4$ showed $p$-type behavior with large $P$ values exceeding the practical level in a wide temperature range of 130–270 K.[8] For such high thermoelectric performances in $M_4$SiTe$_4$ ($M$ = Ta or Nb) below room temperature, the one-dimensional electronic structure with a small band gap most likely plays an important role.[1,6]

The above results suggest that Ta$_4$SiTe$_4$ and Nb$_4$SiTe$_4$ are promising thermoelectric materials for low temperature applications, such as the local cooling of cryogenic electronic devices and cryogenic power generation. However, in order for these compounds to be considered practical materials, it is necessary to prepare a solid solution of Ta$_4$SiTe$_4$ and Nb$_4$SiTe$_4$ and to realize the high thermoelectric performance of both the $n$- and $p$-type materials in the solid-solution samples, in terms of the suppression of lattice thermal conductivity and the optimization of the band gap. As seen in Bi$_2$Te$_3$-Sb$_2$Te$_3$-Bi$_2$Se$_3$ solid solutions and Si-Ge alloys,[9-11] all practical materials contain multiple elements of the same family. Such a solid solution of homologous elements reduces the lattice thermal conductivity and optimizes the band gap, thereby improving the thermoelectric properties. In addition, solid-solution formation is an approach for matching the thermal expansion and mechanical properties of $n$- and $p$-type materials in a device.

Almost all $M_4$SiTe$_4$ compounds synthesized so far contain either Ta or Nb atoms.[1,7,8,12-15] (Ta$_{0.5}$Nb$_{0.5}$)$_4$SiTe$_4$ has only been synthesized and its thermoelectric properties have been reported.[6] Here, we report the thermoelectric properties of whisker crystals of solid solutions between Ta$_4$SiTe$_4$ and Nb$_4$SiTe$_4$ and Mo- or Ti-doped (Ta$_{0.5}$Nb$_{0.5}$)$_4$SiTe$_4$. The Ta$_4$SiTe$_4$-Nb$_4$SiTe$_4$ solid solutions showed a systematic change of $S$ between the end members, while their ρ showed a small value similar to that of Nb$_4$SiTe$_4$. Furthermore, both the Mo- and Ti-doped (Ta$_{0.5}$Nb$_{0.5}$)$_4$SiTe$_4$ samples showed $n$- and $p$-type behavior, respectively, with large $P$ values reaching a practical level.

The whisker crystals of (Ta$_{1-x}$Nb$_x$)$_4$SiTe$_4$ (0 ≤ $x$ ≤ 1), (Ta$_{0.5-y/2}$Nb$_{0.5-y/2}$Mo$_y$)$_4$SiTe$_4$ ($y$ ≤ 0.05), and (Ta$_{0.5-z/2}$Nb$_{0.5-z/2}$Ti$_z$)$_4$SiTe$_4$ ($z$ ≤ 0.05) were synthesized by crystal growth in a vapor phase. A stoichiometric amount of elemental powders and 100% excess Si powder were mixed and sealed in an evacuated quartz tube with 10–20 mg of TeCl$_4$ powder. The addition of an excess Si powder prevents spontaneous combustion when opening the quartz tube. The tube was heated to and maintained at 873 K for 24 h, 1423 K for 96 h, and then furnace cooled to room temperature. The whisker crystals were typically several mm in length and several mm in diameter. The formation of solid solutions between Ta$_4$SiTe$_4$ and Nb$_4$SiTe$_4$ was confirmed by powder X-ray diffraction of pulverized whisker crystals using a MiniFlex diffractometer (RIGAKU) and chemical analysis using an energy-dispersive X-ray (EDX) spectrometer (EDAX Genesis). As shown in the inset of Fig. 1(b), the 130 reflection peak was shifted to a higher angle with increasing the Nb content $x$ without increase of the peak width, suggesting that the solid-solution samples are a single phase. Systematic variation of the chemical composition was observed in the EDX measurements, but the results of EDX might not be quantitative due to the low accelerating voltage of a scanning electron microprobe VE-7800 (KEYENCE) and the overlaps of the characteristic X-ray lines. Therefore, in this study, nominal compositions were used to represent the chemical compositions of the whisker crystals. The electronic resistivity and thermoelectric power measurements between 5 and 300 K were performed using a Physical Property Measurement System (Quantum Design). Uncertainty in the electrical resistivity data was largely dominated by uncertainty in the estimation of the cross sections of the whisker crystals. Uncertainty in the diameter measurements by using an optical microscope VHX-1000 (KEYENCE) is up to 10%, indicating that the uncertainty in the evaluated cross section or electrical resistivity is up to 20%, which has little effect on the discussion in this study. The uncertainty in thermoelectric power data is several percent, which is dominated by a temperature change during a measurement and the uncertainty of voltage measurements.

Figures 1(a) and (b) show the temperature dependences of $S$ and $\rho$ of the whisker crystals of (Ta$_{1-x}$Nb$_x$)$_4$SiTe$_4$ ($0 \leq x \leq 1$), respectively, measured along the $c$ axis. Each sample showed negative $S$ over the entire measured temperature range, indicating that the electron carriers are dominant. In each sample, $|S|$ increased with decreasing temperature from room temperature, reaching a maximum value at 150–200 K. Below this temperature, $|S|$ decreased toward zero with decreasing temperature. The $|S|$ value systematically increased from Nb$_4$SiTe$_4$ to Ta$_4$SiTe$_4$. As seen in Fig. 1(b), the $\rho$ of the solid-solution samples is ~1 mΩ cm at room temperature and ~2 mΩ cm at 100–200 K, which are almost the same as those of Ta$_4$SiTe$_4$ and Nb$_4$SiTe$_4$.[1,6] At lower temperatures, the $\rho$ of the solid-solution samples showed almost constant values of 2–3 mΩ cm similar to Nb$_4$SiTe$_4$, although the $\rho$ of Ta$_4$SiTe$_4$ strongly increased below 50 K.

As shown in Fig. 1(c), $P = S^2/\rho$ of the (Ta$_{1-x}$Nb$_x$)$_4$SiTe$_4$

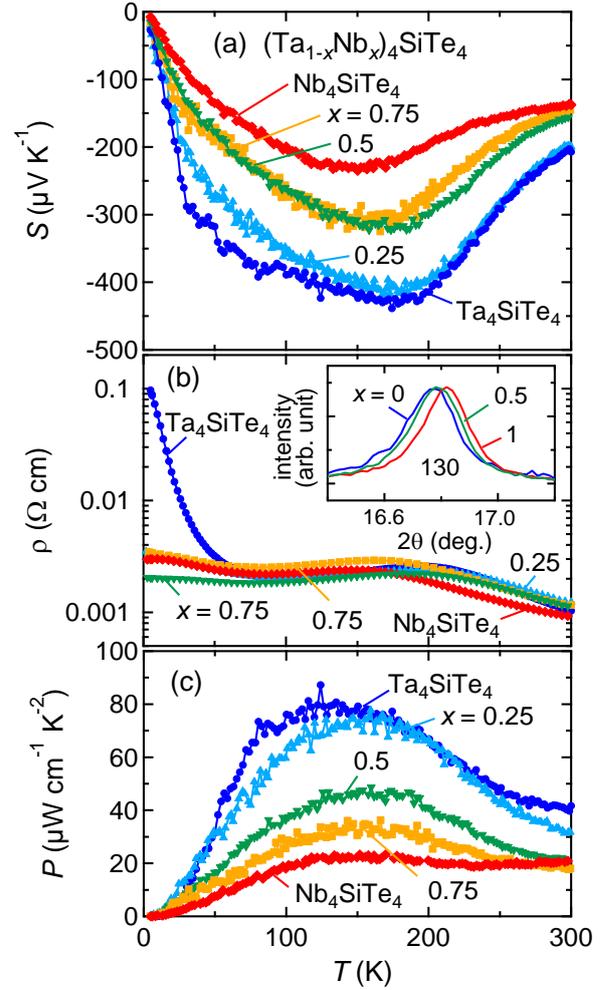

Fig. 1. Temperature dependences of thermoelectric power (a), electrical resistivity (b), and power factor (c) of the whisker crystals of (Ta$_{1-x}$Nb$_x$)$_4$SiTe$_4$ ($0 \leq x \leq 1$). Thermoelectric power and electrical resistivity were measured along the $c$ axis. The inset of (b) shows powder X-ray diffraction patterns around the 130 reflection peaks of pulverized whisker crystals of Ta$_4$SiTe$_4$, (Ta$_{0.5}$Nb$_{0.5}$)$_4$SiTe$_4$, and Nb$_4$SiTe$_4$.

whisker crystals increased with decreasing Nb content $x$. The large $P$ exceeding 35 μW cm$^{-1}$ K$^{-2}$ at room temperature for Bi$_2$Te$_3$-based materials appeared at 130–190 K for $x = 0.5$ and 70–250 K for $x = 0.25$.

Here, we discuss the thermoelectric properties of a series of Mo- or Ti-doped (Ta$_{0.5}$Nb$_{0.5}$)$_4$SiTe$_4$ samples. Figures 2(a) and (b) show the temperature dependences of $S$ and $\rho$ of whisker crystals of Mo-doped (Ta$_{0.5}$Nb$_{0.5}$)$_4$SiTe$_4$, i.e., (Ta$_{0.5-y/2}$Nb$_{0.5-y/2}$Mo$_y$)$_4$SiTe$_4$, respectively. Each sample showed negative $S$ over the entire measured temperature range, indicative of the $n$-type behavior. The peak temperature of $|S|$, which was 180 K in the $y = 0$ sample, increased with increasing Mo content $y$, resulting in $d|S|/dT > 0$ below room temperature at $y = 0.03$. The $|S|$ value decreased with increasing $y$, but the $|S| = 200$ μV K$^{-1}$ of $y = 0.005$ at room temperature was sufficiently large as a thermoelectric material. This Mo doping dependence was similar to the cases of $Ta_4SiTe_4$ and $Nb_4SiTe_4$.[1,6] Mo substitution to the Ta/Nb site represents electron doping, which is expected to increase the electronic carrier concentration.

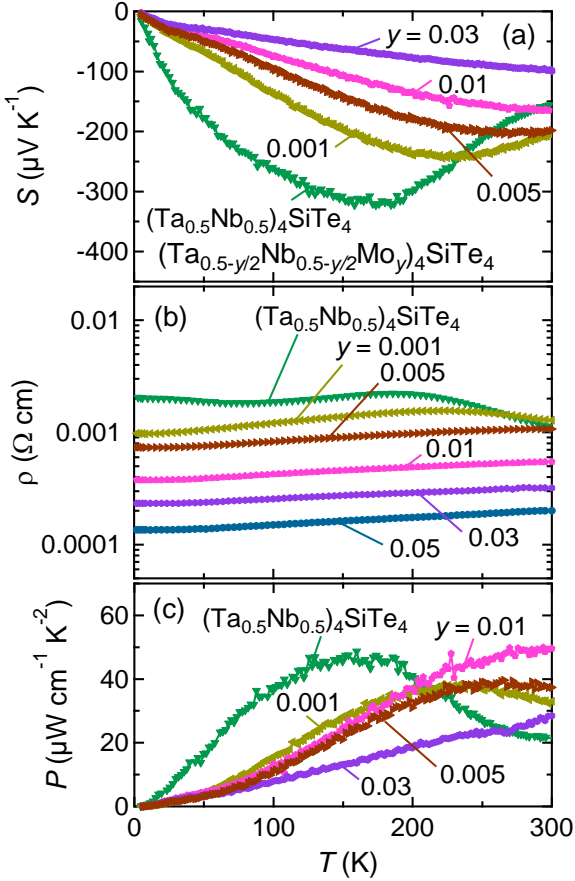

Fig. 2. Temperature dependences of thermoelectric power (a), electrical resistivity (b), and power factor (c) of the whisker crystals of $(Ta_{0.5-y/2}Nb_{0.5-y/2}Mo_y)_4SiTe_4$ ($y \leq 0.05$). Thermoelectric power and electrical resistivity were measured along the $c$ axis.

The electrical resistivity of $(Ta_{0.5-y/2}Nb_{0.5-y/2}Mo_y)_4SiTe_4$ decreased with increasing $y$ and becomes $d\rho/dT > 0$ in the entire measured temperature range at $y \geq 0.005$, reflecting the increase in electron carriers. This behavior is again similar to the cases of Mo-doped $Ta_4SiTe_4$ and $Nb_4SiTe_4$. Figure 2(c) shows the temperature dependence of $P$ of $(Ta_{0.5-y/2}Nb_{0.5-y/2}Mo_y)_4SiTe_4$. The optimum temperature shifted to higher temperature with increasing $y$ and $P$ reached its largest value of 50 μW cm$^{-1}$ K$^{-2}$ in the $y = 0.01$ sample.

Figures 3(a) and (b) show the temperature dependences of $S$ and $\rho$ of the whisker crystals of Ti-doped $(Ta_{0.5}Nb_{0.5})_4SiTe_4$, i.e., $(Ta_{0.5-z/2}Nb_{0.5-z/2}Ti_z)_4SiTe_4$, respectively. At only 0.1% Ti substitution, i.e., $z = 0.001$, a positive $S$ below 230 K was found. The $S$ of $z = 0.001$ was negative at around room temperature, the same as in $(Ta_{0.999}Ti_{0.001})_4SiTe_4$.[8] With decreasing temperature, $S$ increased and showed a maximum value of ~400 μV K$^{-1}$ at 60–150 K, followed by a strong decrease toward zero below 60 K. The maximum $S$ shifted to higher temperatures with increasing Ti content $z$. The temperature dependence of $S$ at $z \geq 0.005$ was similar to that of Ti-doped $Nb_4SiTe_4$.[8] However, unlike the case of $Nb_4SiTe_4$, where the maximum $S$ value decreased to ~100 μV K$^{-1}$ with increasing Ti content, $S$ in the $z = 0.05$ sample retained a large value reaching 150 μV K$^{-1}$.

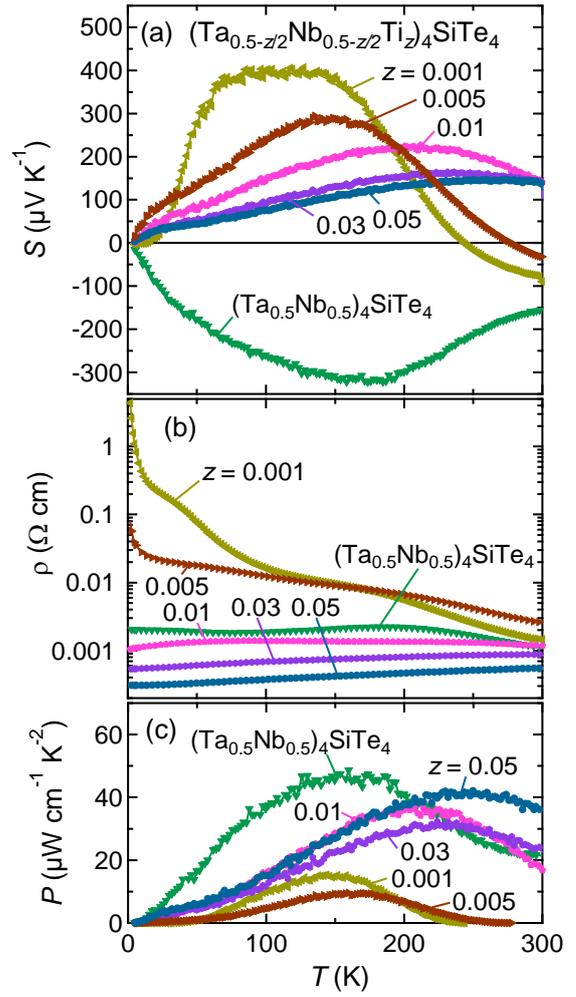

Fig. 3. Temperature dependences of thermoelectric power (a), electrical resistivity (b), and power factor (c) of the whisker crystals of $(Ta_{0.5-z/2}Nb_{0.5-z/2}Ti_z)_4SiTe_4$ ($z \leq 0.05$). Thermoelectric power and electrical resistivity were measured along the $c$ axis. In (c), the power factor in the $p$-type temperature range is shown.

The electrical resistivity of Ti-doped $(Ta_{0.5}Nb_{0.5})_4SiTe_4$ was similar to those of Ti-doped $Ta_4SiTe_4$ in the lightly doped samples and Ti-doped $Nb_4SiTe_4$ in the heavily doped samples. First, $\rho$ of the most lightly doped sample of $z = 0.001$ showed a thermally activated behavior in three temperature regions, similar to $(Ta_{0.999}Ti_{0.001})_4SiTe_4$.[8] As shown in the Arrhenius plot of the electrical conductivity $\sigma$ ($= 1/\rho$) of the $z = 0.001$ sample in Fig. 4, linear behavior appeared at $T < 20$ K ($1000/T > 50$ K$^{-1}$), $T \sim 80$ K ($1000/T \sim 14$ K$^{-1}$), and $T > 200$ K ($1000/T < 5$ K$^{-1}$). The activation energy of each temperature range was estimated to be $E_a = 1.0$, 15, and 75 meV, respectively. As discussed later, these thermally activated behaviors correspond to the thermal excitation of electrons from the donor levels to the conduction band, from the valence band to the acceptor levels introduced by the Ti doping, and across the band gap, respectively. With increasing $z$, these thermally activated behaviors became indistinct, accompanied by the decrease of ρ, resulting in the metallic behavior over the entire measured temperature range in the $z \geq 0.03$ samples.

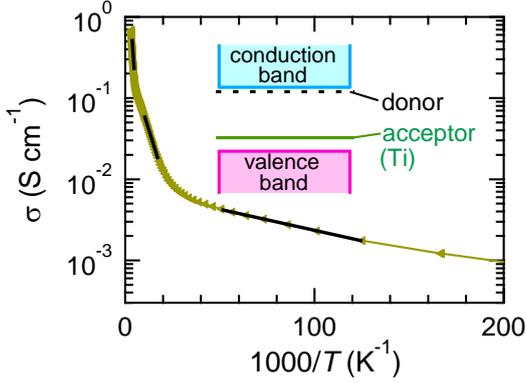

Fig. 4. Arrhenius plot of electrical conductivity of the $(Ta_{0.5-z/2}Nb_{0.5-z/2}Ti_z)_4SiTe_4$ ($z = 0.001$) whisker crystal measured along the $c$ axis. Solid lines represent the results of linear fits to 213–273 K, 58–100 K, and 8–20 K data. The inset shows a schematic energy diagram of Ti-doped $(Ta_{0.5}Nb_{0.5})_4SiTe_4$.

Figure 3(c) shows the $P$ of $(Ta_{0.5-z/2}Nb_{0.5-z/2}Ti_z)_4SiTe_4$ in the $p$-type temperature region. The $z \leq 0.005$ samples showed large $S$ and ρ, yielding the $P$ values of less than 20 μW cm$^{-1}$ K$^{-2}$. With increasing $z$, $P$ increased and reached 40 μW cm$^{-1}$ K$^{-2}$ at 240 K in the $z = 0.05$ sample. This $P$ was smaller than the largest value for $p$-type $M_4SiTe_4$ ($P = 57$ μW cm$^{-1}$ K$^{-2}$ for $(Nb_{0.95}Ti_{0.05})_4SiTe_4$), but was comparable to the room-temperature value of $Bi_2Te_3$-based materials.

One of the purposes of synthesizing solid solutions in the study of thermoelectric materials is the optimization of thermoelectric properties by controlling the size of the band gap. Here we considered this by comparing the σ of the $z = 0.001$ sample, which showed the most pronounced thermally activated behavior in this study, with those of Ti-doped $Ta_4SiTe_4$ and $Nb_4SiTe_4$. As shown in Fig. 4, σ of the $z = 0.001$ sample showed thermally activated behavior at $T < 20$ K, $T \sim 80$ K, and $T > 200$ K with $E_a = 1.0$, 15, and 75 meV, respectively, and the third one corresponded to the thermal excitation of electrons across the band gap. Similar thermal excitations with $E_a = 96$ and 57 meV were observed in $(Ta_{0.999}Ti_{0.001})_4SiTe_4$ and $(Nb_{0.998}Ti_{0.002})_4SiTe_4$, respectively.[8] The observed $E_a = 75$ meV in the $z = 0.001$ sample was an intermediate value between 96 and 57 meV, demonstrating that the band gap in $M_4SiTe_4$ was controlled by the formation of the solid solution. In addition, the thermally activated behavior at $T \sim 80$ K with $E_a = 15$ meV corresponded to the thermal excitation of electrons from the valence band to the acceptor levels introduced by Ti doping. This $E_a$ value was smaller than $E_a = 37$ meV in $(Ta_{0.999}Ti_{0.001})_4SiTe_4$.[8] Ti-doped $(Ta_{0.5}Nb_{0.5})_4SiTe_4$ samples showed positive $S$ above the lower temperatures than those in Ti-doped $Ta_4SiTe_4$, corresponding to the fact that the doped Ti atoms in the solid-solution samples acted as acceptors at lower temperature.

Finally, we discuss the effects of solid-solution formation on thermoelectric properties of $M_4SiTe_4$. First, $|S|$ of $(Ta_{1-x}Nb_x)_4SiTe_4$ simply increased from $Nb_4SiTe_4$ to $Ta_4SiTe_4$, as shown in Fig. 1(a). A similar variation appeared in the Mo- or Ti-doped solid-solution samples. For example, the maximum $S$ of Ti-doped $(Ta_{0.5}Nb_{0.5})_4SiTe_4$ was ~400 μV K$^{-1}$, which is an intermediate value of ~250 μV K$^{-1}$ for Ti-doped $Nb_4SiTe_4$ and ~600 μV K$^{-1}$ for Ti-doped $Ta_4SiTe_4$.[8] These results suggested that the density of states near the band edges of conduction and valence bands continuously increased from $Nb_4SiTe_4$ to $Ta_4SiTe_4$. In contrast, ρ of $(Ta_{1-x}Nb_x)_4SiTe_4$ was unexpectedly low at low temperatures. Generally, the ρ of solid-solution samples is expected to be larger than those of the end members due to the chemical disorder, particularly in the one-dimensional case. As seen in Fig. 1(b), ρ of the $(Ta_{1-x}Nb_x)_4SiTe_4$ samples not only showed similar values to those of the end members above 50 K but was also much smaller than that of $Ta_4SiTe_4$ below this temperature. Similar behavior appeared also in the Ti-doped samples. These results were surprising, because $M_4SiTe_4$ has a one-dimensional electronic structure.[1,6] At present, the mechanism of the suppression of ρ is unclear, but it would be interesting if the Dirac-like band dispersion in $M_4SiTe_4$, where the valence and conduction bands cross at around the Fermi energy, plays an important role.[1,6] In addition to the $S$ and ρ, thermal conductivity κ is expected to be affected by the solid-solution formation. The phonon contribution of κ of the solid-solution samples is expected to be considerably smaller than those of the end members, such as in Si-Ge and $Mg_2Si$-$Mg_2Ge$ alloys.[10,11,16] At present, it is not possible to evaluate the κ of solid-solution samples, because the prepared whisker crystals are too thin. However, the above discussion on the ρ suggested that the phonon contribution of κ of $M_4SiTe_4$ could be reduced by making the solid solution without increase of ρ, giving rise to a large figure of merit $Z = S^2/\rho\kappa$ below room temperature.

In summary, we have reported on the synthesis and thermoelectric properties of the whisker crystals of $(Ta_{1-x}Nb_x)_4SiTe_4$ and Mo- or Ti-doped $(Ta_{0.5}Nb_{0.5})_4SiTe_4$. $(Ta_{1-x}Nb_x)_4SiTe_4$ and Mo-doped $(Ta_{0.5}Nb_{0.5})_4SiTe_4$ samples showed negative thermoelectric power, while positive thermoelectric power appeared in the Ti-doped (Ta$_{0.5}$Nb$_{0.5}$)$_4$SiTe$_4$ samples. Both the *n*- and *p*-type samples showed large power factors exceeding the room-temperature values of Bi$_2$Te$_3$-based materials below room temperature. The thermoelectric power of (Ta$_{1-x}$Nb$_x$)$_4$SiTe$_4$ samples systematically increased from Nb$_4$SiTe$_4$ to Ta$_4$SiTe$_4$. In contrast, their electrical resistivity was small and comparable to that of Nb$_4$SiTe$_4$, which was surprising because *M*$_4$SiTe$_4$ was reported to have a strongly one-dimensional electronic structure. These results indicated that *M*$_4$SiTe$_4$ is promising as a practical material for low temperature applications.

**Acknowledgments**

The authors are grateful to F. Matsunaga for his help with the experiments. This work was partly carried out under the Visiting Researcher Program of the Institute for Solid State Physics, the University of Tokyo and supported by the Collaborative Research Project of Materials and Structures Laboratory, Tokyo Institute of Technology, JSPS KAKENHI (Grant Numbers: 19K21846, 19H05823, and 20H02603), the Asahi Glass Formation, the Research Foundation for the Electrotechnology of Chubu, and the Inamori Foundation.